\documentclass[webpdf,modern,medium]{oup-authoring-template}

\onecolumn 

\usepackage{endfloat}
\usepackage{bm}
\usepackage{tabularx}
\usepackage{booktabs}

\graphicspath{{Fig/}}

\theoremstyle{thmstyleone}%
\theoremstyle{thmstyletwo}%
\theoremstyle{thmstylethree}%

\begin{document}

\journaltitle{Journal Title Here}
\DOI{DOI added during production}
\copyrightyear{YEAR}
\pubyear{YEAR}
\vol{XX}
\issue{x}
\access{Published: Date added during production}
\appnotes{Paper}

\firstpage{1}


\title[Complete cancer prevalence predictions]{Complete Cancer Prevalence Predictions via LASSO Regularization and Flexible Parametric Survival Cure Models}

\author[1,2$\ast$]{Fabrizio Di Mari\ORCID{0009-0004-7779-7297}}
\author[3]{Roberta De Angelis\ORCID{0000-0001-7787-3043}}
\author[1]{Therese ML Andersson\ORCID{0000-0001-8644-9041}}
\author[4]{Enoch Yi-Tung Chen\ORCID{0000-0003-2448-708X}}
\author[3]{Silvia Rossi\ORCID{0000-0002-9815-2904}}
\author[1]{Paul W Dickman\ORCID{0000-0002-5788-3380}}
\author[5]{Roberto Rocci\ORCID{0000-0003-4366-7975}}
\author[1]{Mark Clements\ORCID{0000-0003-4518-5670}}

\address[1]{\orgdiv{Department of Medical Epidemiology and Biostatistics}, \orgname{Karolinska Institutet}, \orgaddress{\street{Nobels väg 12A}, \postcode{171 65}, \state{Solna}, \country{Sweden}}}
\address[2]{\orgdiv{Department of Economics, Statistics and Finance}, \orgname{University of Calabria}, \orgaddress{\street{Via Pietro Bucci}, \postcode{87036}, \state{Rende CS}, \country{Italy}}}
\address[3]{\orgdiv{Department of Oncology and Molecular Medicine}, \orgname{Italian National Institute of Health}, \orgaddress{\street{Viale Regina Elena 299}, \postcode{00161}, \state{Rome}, \country{Italy}}}
\address[4]{\orgdiv{Department of Statistical Science}, \orgname{Sapienza University of Rome}, \orgaddress{\street{Piazzale Aldo Moro 5}, \postcode{00185}, \state{Rome}, \country{Italy}}}

\corresp[$\ast$]{Corresponding author. \href{email:fabrizio.dimari@unical.it}{fabrizio.di.mari@ki.se}}

\received{Date}{0}{Year}
\revised{Date}{0}{Year}
\accepted{Date}{0}{Year}



\abstract{Cancer prevalence is a key measure of disease burden, and accurate projections are essential for healthcare planning and resource allocation. The Prevalence Incidence Analysis Model (PIAMOD) is widely used to estimate and project cancer prevalence in population-based studies by combining incidence rates and net survival within a forward calculation framework. We propose an improved approach to projecting cancer prevalence by using regularization techniques to select the optimal incidence model and flexible parametric survival cure models to improve the estimation of net survival. The proposed and conventional approaches were compared through a simulation study and an application to population-based data on patients diagnosed with colon cancer in Sweden between 1958 and 2019. Applying the proposed methodology, colon cancer prevalence in Sweden was projected through 2030. The number of individuals living with a previous diagnosis of colon cancer is projected to continue increasing, reaching approximately 550 per 100\,000 males and 500 per 100\,000 females by 2030. Careful model selection is essential when estimating measures of disease burden such as cancer prevalence. The projected increase in colon cancer prevalence in Sweden reflects continued improvements in survival for both sexes and highlights the growing number of individuals living with colon cancer.} 


\keywords[Abbreviations]{Prevalence Incidence Analysis MODel (PIAMOD), age-period-cohort (APC), likelihood ratio test (LRT), Weibull mixture cure models (WMCM), flexible parametric survival cure models (FPSCM), mean squared error (MSE), Least Absolute Shrinkage and Selection Operator (LASSO)}

\maketitle


\section{Introduction}
\label{sec:intro}
Prevalence is one of the primary measures used to assess the impact of a disease on a population. It can be defined in several ways, depending on the disease of interest. In the cancer context, it is generally defined as the proportion of individuals who have been diagnosed with cancer within a defined population and are alive at a specific time. There are two major methodologies to estimate cancer prevalence: \textit{Direct Methods} and \textit{Indirect Methods}. 

Direct methods \citep{feldman1986,coldman1992} use individual incidence and follow-up data from population-based cancer registries to compute \textit{limited duration prevalence} by counting the number of patients alive in the population at a specific time point. They are distinguished by whether cure time points are assumed and how lost-to-follow-up cases are handled. These methods inherently underestimate prevalence, particularly when the registry is newly established and has very limited follow-up. Individuals diagnosed before the establishment of the registry who are still alive at the time of estimation are not included in the analysis. The size of this bias depends on the length of the observation period. \cite{capocaccia1997} addressed this drawback by proposing the \textit{completeness index method}, which models both the observed and unobserved parts of prevalence to derive total prevalence estimates from partial measures. However, these methods provide prevalence only within the observation period and do not estimate a model able to predict it beyond the observed calendar years. Recently, \cite{deangelis2024} projected cancer prevalence in Europe using linear regression after applying the completeness index method. Their comprehensive study covered 29 European countries and estimated prevalence in 2020 for 32 malignant tumours. While their approach could provide reasonable results for short-term projections, more flexible models are required to help reduce potential bias from overly simplistic models when extrapolating further into the future.

Indirect approaches use incidence, mortality, and follow-up data to calculate cancer prevalence through statistical modelling. The \textit{Mortality Incidence Analysis MODel} (MIAMOD) and the \textit{Prevalence Incidence Analysis MODel} (PIAMOD) \citep{verdecchia1989, deangelis1994, mariotto2000, verdecchia2002} have been widely used to estimate and project the prevalence (and mortality) of cancer, as well as its economic burden in various countries \citep{mariotto2006,rossi2013,mariotto2020}. 

The PIAMOD method estimates the prevalence of cancer under the assumption of irreversible chronic disease. It is typically applied in population-based studies using data from cancer registries and other official statistics. It requires modelling two estimands: incidence rate and net survival. These two quantities are combined to compute prevalence through a forward-type calculation. Incident cases are specified as an age-period-cohort (APC) model using Poisson regression, where the linear predictor is either a polynomial or a spline function of the variables \textit{age}, \textit{calendar period}, and \textit{cohort}. A stepwise procedure \citep{hocking1976} based on the likelihood ratio test (LRT) is used to either select the degrees of the age, period and cohort effects in the polynomial or the degrees of freedom of the restricted cubic spline functions. 

Net survival, the survival as if the disease of interest were the only possible cause of death, is estimated within a relative survival framework \citep{ederer1961}. In this setting, the all-causes hazard of patients with the disease of interest is decomposed as the sum of the hazard due to the disease-specific cause of death and the hazard due to all other causes. The latter hazard is estimated from routine data sources (e.g. national or regional life tables), under the assumption that the disease is \textit{rare} in the general population, namely that removing deaths due to the specific cause from these sources have a negligible affect on the estimated all-cause mortality rates. The validity of this assumption, also known as the \textit{exchangeability assumption} \citep{eloranta2020}, has been explored by \cite{talback2011} in the context of population-based cancer survival studies. The second required assumption is the \textit{independent risks} assumption, namely that the risk of dying from the disease is independent of the risk of dying from other causes. These assumptions are crucial in population-based studies, and they imply that the estimand we are targeting is net survival \citep{dimari2026}. In the original PIAMOD paper \citep{verdecchia2002}, the authors proposed estimating net survival in two ways. The first approach is to use non-parametric estimators, such as the Hakulinen estimator \citep{hakulinen1985}, the Ederer I and II estimators \citep{ederer1961}, or the later proposed Pohar-Perme estimator \citep{perme2012} which has been shown, under non-informative censoring, to a be consistent estimator of net survival. The second is to model net survival semi-parametrically using Weibull mixture cure models (WMCM), originally proposed by \cite{deangelis1999}, and later extended by \cite{lambert2007}. 

Both the classical APC model selection and the net survival modelling approach have several drawbacks. The APC model selection procedure does not consider any measure of out-of-sample performance. When projecting prevalence into the future, the primary concern should be selecting a model that can generalize to data outside the observed period. \cite{novikov2020} proposed a method for selecting the APC model in the PIAMOD framework by measuring the error between the estimated prevalence and observed limited duration prevalence. They used the last observed 10 years to select the model and validated it against the standard LRT driven approach. This method is computationally inefficient, as it requires calculating prevalence for all the APC models considered. This process becomes cumbersome when comparing not only numerous APC polynomials (or spline functions), but also different net survival estimators. Additionally, model selection based on limited duration prevalence is biased by the incomplete observation time of prevalence, and the extent of this bias depends on the length of the observation period. As a result, this method cannot be applied to populations covered by recent established cancer registries. With regard to net survival, neither the Pohar-Perme estimator nor the WMCM are fully suitable for extrapolation. On one hand, the Pohar-Perme estimator requires projecting constant net survival beyond the last observed calendar year, as is the case with all other non-parametric estimators. Moreover, it is highly variable at the tail of the survival curve, particularly for older age groups \citep{lambert2015}. On the other hand, several studies have shown that the WMCM may lack sufficient flexibility to fit the observed data and capture the fraction of cured individuals \citep{andersson2011,yu2013}. 

To improve the PIAMOD procedure, we propose a Least Absolute Shrinkage and Selection Operator (LASSO) \citep{tibshirani1996} driven model selection procedure similar to the two-step estimator proposed by \cite{dimari2023} in the context of penalized clusterwise linear regression. However, instead of minimizing the Bayesian Information Criterion (BIC) \citep{schwarz1978} on the LASSO solutions, the strength of the penalization parameter is chosen to minimize the prediction error of incident cases in the last observed, say \textit{h}, years, which are left out during the training phase. In this way, we adopt a local approach by giving a higher weight to observed incidence time trends in the most recent years. To mitigate the arbitrariness in the choice of \textit{h} we build an ensemble of model selectors by varying the test set dimension from 1 year to $h_{\text{max}}$. We then select the powers of the APC polynomial that appear at least 50\% of the times across all decisions and refit the model on the whole dataset with the selected powers. 

Regarding net survival estimation, we use flexible parametric survival cure models (FPSCM) originally proposed by \cite{andersson2011}, and later extended by \cite{jakobsen2020}, aiming to reduce model misspecification that may arise from using too rigid models, while also benefiting from the lower variability associated with parametric approaches, especially at the tails of the survival curve. 

The structure of the paper is as follows. In Section \ref{sec:model}, we introduce the general model, derive the estimands of interest, and describe our proposed estimation procedures. In Section \ref{sec:sim}, we conduct an extensive simulation study to compare the performance of our LASSO-driven model selection algorithm with that of the stepwise selection procedure based on LRTs. In Section \ref{sec:appl}, we illustrate how this machinery can be used to provide insights into the prevalence of colon cancer in Sweden by analysing data from 1958 to 2019 and projecting prevalence up to 2030. Finally, in Section \ref{sec:conc}, we discuss the results and their implications.

\section{Modelling framework and estimands of interest}
\label{sec:model}
\begin{figure}[!ht]
    \centering  
    \resizebox{\textwidth}{!}{
    \includegraphics{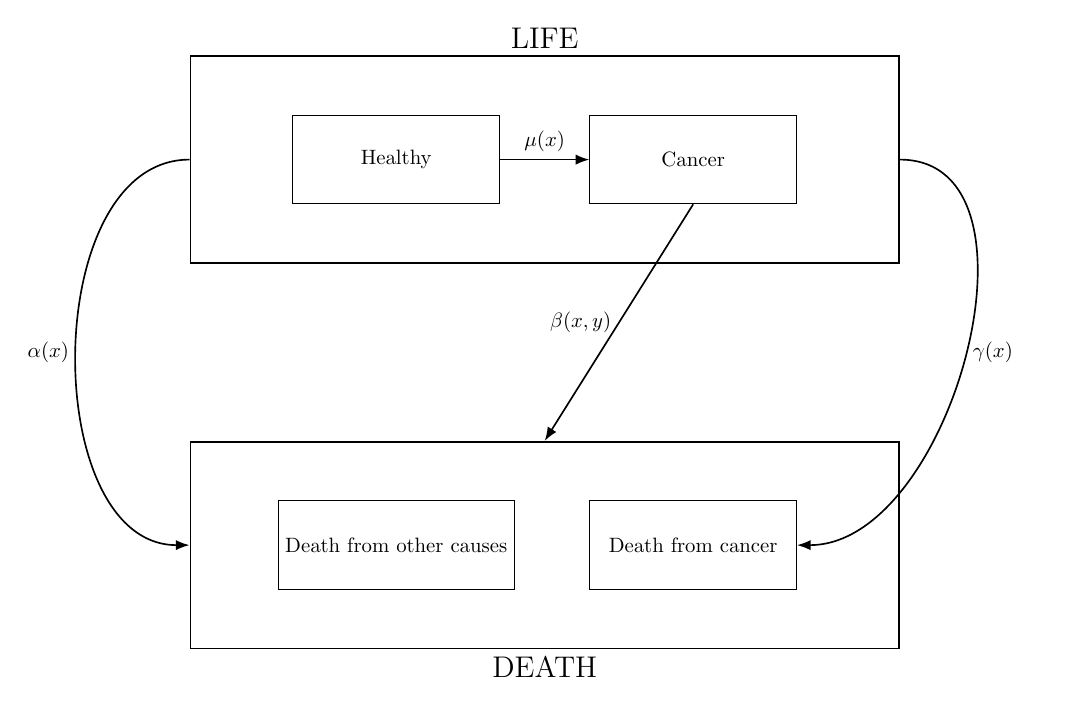}
    }
    \caption{Representative diagram of the irreversible illness-death model.}
    \label{fig:compartimental}
\end{figure}

Suppose an irreversible illness-death model is used to describe the natural history of cancer, depicted in Figure \ref{fig:compartimental} and adapted from \citet{verdecchia1989}.
There are two transient states, healthy and living with cancer, and two absorbing states, death due to cancer and death due to other causes. As shown in \citet{verdecchia1989}, only the transitions of interest are represented, and the backward link from cancer to healthy is deliberately omitted. In other words, cancer is assumed to be an irreversible disease. The model framework and the estimands of interest are derived following the reasoning in \citet{verdecchia1989,verdecchia2002}.

Consider the disease process affecting individuals in a defined birth cohort. Let $\mu(x)$ represent the hazard of getting cancer for healthy individuals at age $x$, and let $\nu(x)$ denote the probability of being in the sick state at age $x$, conditional on having survived up to age $x$. Let $\alpha(x)$ be the death hazard due to all-cause deaths at age $x$, and $\gamma(x)$ the hazard due to cancer-specific death at age $x$, irrespective of the age at diagnosis. Furthermore, let $\beta(x,y)$ represent the all-cause death hazard for individuals at age $x$ who get diagnosed with cancer at age $y$. We are interested in expressing the prevalence of the disease $\nu(x)$ in the cohort at age $x$ as functions of $\mu(x)$, $\beta(x,y)$ and $\alpha(x)$.

The all-cause survival function $S(x)$ for any individual in the cohort is given by
\begin{equation*}
    S(x) = \exp\left\{ - \int_{0}^{x} \alpha(u) \text{d}u \right\}.
\end{equation*}
Then, the probability of being in the sick state at age $x$ for people in the cohort can be written as
\begin{equation*}
    S(x)\nu(x)=\int_{0}^{x}S(\tau)\{1-\nu(\tau)\}\mu(\tau)\exp\left\{ - \int_{\tau}^{x} \beta(u,\tau) \text{d}u \right\} \text{d}\tau,
\end{equation*}
from which we can derive a Volterra integral Equation of the second kind for prevalence \citep{linz1969}:
\begin{equation}
\label{eq:prevalence}
    \nu(x)=\int_{0}^{x}\{1-\nu(\tau)\}\mu(\tau)\exp\left[ - \int_{\tau}^{x} \{\beta(u,\tau)-\alpha(u)\}\text{d}u \right] \text{d}\tau,
\end{equation}
namely the probability of being sick at age $x$, conditional on having survived up to age $x$, is given by the probability of not being sick prior to age $\tau$, multiplied by the instantaneous probability of getting sick at age $\tau$, multiplied by the probability of surviving the extra death risk between ages $\tau$ and $x$, integrated over all ages $\tau \leq x$ at which the individual could have been diagnosed with cancer.

In practice, the quantity of interest we have derived in continuous time reduces to discrete time. To derive Equation (\ref{eq:prevalence}) in discrete time, we assume that diagnosis and death can only occur at the middle point between two consecutive birthdays. $\mu_j$ is now the probability of being diagnosed with cancer between ages $j$ and $j+1$ for healthy individuals at exact age $j$. Furthermore, we define the discrete probability of surviving the extra death risk of dying of cancer from age $j$ to age $i$, denoted by $S_{E,ij}$, namely net survival, as follows
\begin{equation*}
    S_{E,ij} = \exp\left\{ - \sum_{\ell=j}^{i} ( \beta_{\ell j} -\alpha_\ell ) \right\},
\end{equation*}
where $\beta_{\ell j}$ is the probability of dying from all causes between age $\ell$ and $\ell+1$ for an individual diagnosed with cancer between age $j$ and $j+1$, and $\alpha_\ell$ is the probability of dying from any cause between ages $\ell$ and $\ell+1$. We can now rewrite Equation (\ref{eq:prevalence}) in discrete terms as follows
\begin{equation}
    \label{eq:prevalencefinal}
    \nu_i = \sum_{j=0}^{i} (1-\nu_j)\mu_jS_{E,ij}.
\end{equation} 
Equation (\ref{eq:prevalencefinal}) provides the age-specific prevalence for a given birth cohort, assuming that both incidence and net survival are known and $\nu_0=0$, that is, zero at age zero. By considering all birth cohorts within the observation period, we can reconstruct the complete cancer prevalence sequence for the entire observation period. Given a calendar year $t$, the average prevalence proportion at age $i$ during year t, denoted by $N_i(t)$, can be computed as follows
\begin{equation}
\label{eq:avgprev}
    N_i(t) = \left( \frac{\nu_{i,z} + \nu_{i+1,z-1} }{2} \right),
\end{equation}
where $z=t-i$ is the birth cohort. Equation (\ref{eq:avgprev}) follows from the assumption that diagnoses and deaths can only occur at the middle point between two consecutive birthdays.

\subsection{Modelling incidence: the age-period-cohort model}
\label{sec:incidence}
Let $I$ be a maximum age and $T$ the number of years in the observation period. Let $m_{it}$ be the number of new cancer cases at age $i$, $i = 1,\dots,I$, and year $t$, $t = 1,\dots,T$. We assume independent $m_{it} | P_{it} \sim \text{Poisson}(\mu_{it}P_{it})$, where $P_{it}$ represents the size of the cohort at age $i$ and year $t$ and $\mu_{it}$ is the corresponding incidence rate. We assume the incidence rate to be a polynomial function of age, period and cohort in the logit scale
\begin{equation}
    \label{eq:incidence}
    \text{logit}(\mu_{it}) = a_0 + \sum_{k=1}^{K_1}a_k \, i^k + \sum_{k=2}^{K_2}b_k \, t^k + \sum_{k=1}^{K_3}c_k \, (t-i)^k.
\end{equation}
This allows us to take into account independent effect of age, period and cohort in the incidence trend. 
Note that in APC models, the linear components of age, period, and cohort are not jointly identifiable due to the exact linear dependency (\textit{cohort} = \textit{period} - \textit{age}). To ensure model identifiability, we exclude the linear term for period, which avoids collinearity with the linear components of age and cohort. This specific parametrization, which suppresses the linear effect of period, is particularly meaningful when deriving prevalence projections (see Section \ref{sec:projections}).

The regression parameters $\bm{\theta}=[a_0,...,a_{K_1},b_2,...,b_{K_2},c_1,...,c_{K_3}]$ are estimated by maximizing the log likelihood function, which can be written as follows
\begin{equation}
\label{eq:apcllik}
    \ell(\bm{\theta}) = \sum_{i,t} \left[m_{it}\log\{P_{it}\mu_{it}(\bm{\theta})\} - P_{it}\mu_{it}(\bm{\theta})\right].
\end{equation}

In the original model proposal \citep{verdecchia1989}, the degrees of the polynomial $K_1$, $K_2$ and $K_3$ are chosen by a forward stepwise procedure, adding coefficients based on the outcomes of the LRT. However we aim to select a model that can be generalized well beyond the observation period $\{1,\dots,T\}$, and we will do this by estimating the APC trends locally, considering the most recent calendar years.

\subsubsection{LASSO-driven APC model selection procedure}
\label{sec:lassoselection}
Suppose we are interested in basing our predictions on the best fit to the last $h$ years of observations. Let us consider a modification of Equation (\ref{eq:apcllik}) that includes an $L_1$ norm penalty on the parameters, and considering only the first $T-h$ years in our training set:
\begin{equation}
    \label{eq:apcpenllik}
    \sum_{t=1}^{T-h}\sum_i\left[m_{it}\log\{P_{it}\mu_{it}(\bm{\theta})\} - P_{it}\mu_{it}(\bm{\theta})\right] - \lambda||\bm{\theta}||_1.
\end{equation}
From Equation (\ref{eq:apcpenllik}) we compute the full LASSO path as shown in \citet{park2007}. This procedure returns a sequence of $L+1$ knots of the LASSO solution, $\{ \bm{\tilde{\theta}}_l\}_{l=0}^{L}$, where each $\bm{\tilde{\theta}}_l$ has some values shrunk to 0 based on the value of the corresponding $\lambda$ in the penalty term of Equation (\ref{eq:apcpenllik}). As a result, for each $\bm{\tilde{\theta}}_l$ there exist $\mathcal{J}_{l} \subseteq \{0,...,K\}=\mathcal{J}$, with $K=K_1+(K_2-1)+K_3$,  such that $|\tilde{\theta}_{l,j}| \neq 0 \; \; \forall j \in \mathcal{J}_{l}$ and $|\tilde{\theta}_{l,j}| = 0 \; \forall j \not\in \mathcal{J}_{l}$. Then, we set $\lambda=0$ in Equation (\ref{eq:apcpenllik}) and fit the model only on the powers associated to the non-zero coefficient of the index set $\mathcal{J}_{l}$ to obtain the unpenalized solution $\bm{\hat{\theta}}_l$. Finally, we compute the Mean Squared Error (MSE) on the last $h$ years which we have left out until now
\begin{equation}
\label{eq:mse}
    \text{MSE}(\bm{\hat{\theta}}_l) = \frac{1}{h}\sum_{t=T-h+1}^{T} \left\{ \sum_i m_{it} - \sum_i P_{it} \cdot \mu_{it}(\bm{\hat{\theta}}_l) \right\}^2.
\end{equation}
The powers which we then select are the ones in the index set $\mathcal{J}_{l^*}$, with $\displaystyle l^*=\operatorname*{argmin}_{0 \leq l \leq L} \text{MSE}(\bm{\hat{\theta}}_l)$.

The choice of $h$ however is somewhat arbitrary, and a different choice of $h$ leads to different selected powers. Therefore, we mitigate the effect of the choice of $h$ by repeating the procedure for $h=1,...,h_{\text{max}}$, and then selecting the powers which appears at least 50\% of the times across all model selection algorithms in an ensamble fashion \citep{zhou2025}. This ensures that the powers we use for predictions are stable, and they do not depend heavily on the choice of $h$. The pseudocode for the model selection procedure can be summarized in Algorithm \ref{alg:apcsel}.
\begin{algorithm}
\caption{LASSO-driven APC model selection}
\label{alg:apcsel}
\begin{algorithmic}[1]
    \For{$h=1$ to $h_{\max}$}
        \State Split the dataset into training and test sets according to $h$.
        \State Compute the $L+1$ knots of the LASSO path $\{\tilde{\boldsymbol{\theta}}_{l}^{(h)}\}_{l=0}^{L}$ on the training set.
        \For{$l=0$ to $L$}
            \State Fit the unpenalized model~(\ref{eq:apcpenllik}) on the active set $\mathcal{J}_{l}^{(h)}$ of $\tilde{\boldsymbol{\theta}}_{l}^{(h)}$ to obtain $\hat{\boldsymbol{\theta}}_{l}^{(h)}$.
            \State Compute $\operatorname{MSE}(\hat{\boldsymbol{\theta}}_{l}^{(h)})$ as in Equation~(\ref{eq:mse}).
        \EndFor
        \State Select the active set $\mathcal{J}_{l_h^*}^{(h)}$ such that $\displaystyle l_h^*=\operatorname*{arg\,min}_{0\leq l \leq L}\operatorname{MSE}(\hat{\boldsymbol{\theta}}_{l}^{(h)})$.
    \EndFor
    \State Select the powers that appear in at least $50\%$ of the sets $\{\mathcal{J}_{l_h^*}^{(h)}\}_{h=1}^{h_{\max}}$.
\end{algorithmic}
\end{algorithm}
Once the model is selected, we refit the (unpenalized) model on the whole observation period with the selected powers.

\subsection{Modelling net survival: flexible parametric survival mixture cure models}
\label{sec:netsurvival}
Let $\bm{z} \in \mathbb{R}^{q+1}$ be a vector of covariates, in our case \textit{sex, age, calendar period}, possibly some interactions between the variables, and first entry equal to 1, and let $S_E(t,\bm{z})$ be net survival after $t$ years from diagnosis. Then, we write $S_E(t,\bm{z})$ as follows
\begin{equation*}
     S_E(t,\bm{z})=\pi(\bm{z})+(1-\pi(\bm{z}))S_{E_1}(t,\bm{z}).
\end{equation*}
The mixing proportion $\pi$ modulates the cure fraction in the patients population (the net survival associated with it is identically equal to 1) and is linked to a linear predictor on the covariates $\bm{z}$ through a logit link
\begin{equation*}
    \text{logit}(\pi(\bm{z}))=\bm{z}^{\text{T}}\bm{\bm{\beta}},
\end{equation*}
while the survival of the uncured patients $S_{E_1}(t,\bm{z})$ is specified as follows
\begin{equation}
\label{eq:survuncured}
    g(S_{E_1}(t,\bm{z}))= \eta(t,\bm{z},\bm{\phi}) = s_0(t,\gamma_0)+\bm{z}^{\text{T}}\bm{\xi}.
\end{equation}
The link function $g$ is chosen to model the log cumulative (excess) hazard function 
\begin{equation*}
    g(\cdot) = \log(-\log(\cdot)),
\end{equation*}
while the time effect in Equation (\ref{eq:survuncured}) is a restricted cubic spline function of logarithm of time to allow for more flexibility in the estimation of the baseline (excess) hazard of the uncured group. The regression parameters $\bm{\beta}$ and $\bm{\phi}=(\bm{\gamma}_0,\bm{\xi})$ are estimated through maximum likelihood estimation based on individual follow-up data as shown in \citet{jakobsen2020}.

\subsection{Deriving prevalence projections}

\label{sec:projections}
Projecting cancer prevalence beyond the observation period requires some working assumptions, on one hand in both incidence rates and net survival extrapolation, and on the other hand on the population growth. These assumptions are the same as those described in the original PIAMOD paper \citep{verdecchia2002} and are  summarized below for completeness.

Projections of modelled incidence rates into future calendar years can be made by assuming that the age and cohort effects observed during the study period continue unchanged. This approach is generally appropriate for cancer.
In contrast, period effects, which represent shifts in incidence rates affecting all age groups simultaneously, may have been observed historically but are not assumed to persist in future projections. As a result, they are excluded from forward projections. Instead, only the drift component associated with the linear trend of the cohort, which is indistinguishable from a linear period trend, is extended into the future. In this context, the use of the logit link function in Section \ref{sec:incidence}, rather than the logarithmic link, is particularly appropriate, as the latter could otherwise lead to unrealistic exponential growth of incident cases over time.

To project net survival, two approaches can be considered. A conservative scenario assumes that cancer patients' survival remains constant during the projection period, that is no further improvements are expected beyond those observed in past years. Conversely, an optimistic scenario assumes that survival continues to improve at the same rate observed in recent years. The most appropriate choice is not to adopt one hypothesis over the other, but to use both in order to provide a plausible range for the projected rates.

For the assumptions concerning population projections, the number of births and age-specific mortality for causes other than the cancer of interest are assumed constant over the projection period, based on the most recent available data. 
Final population projections are derived for each age group by identifying the corresponding birth cohorts and adjusting for their expected mortality.

Finally, projected prevalence patterns are reconstructed using the projected incidence rates and survival estimates using Equations (\ref{eq:prevalencefinal}) and (\ref{eq:avgprev}).

\section{Simulation study}
\label{sec:sim}
In this section, we followed \citet{morris2019} framework to design a simulation study for evaluating, on the one hand, the generalization performance of the incidence rate estimator obtained through our proposed model selection algorithm and, on the other hand, whether the algorithm is able to select the powers of the polynomial regression described in Section \ref{sec:incidence} which defines the true data-generating process. These objectives were also explored to evaluate the sensitivity of the proposed procedure with respect to the choice of the hyper-parameter $h_{max}$. 

We simulated incidence rates from Equation (\ref{eq:incidence}), using different parameters while assuming the population remained constant over time as if we were estimating standardized incidence rates. The at risk population was the 2013 European Standard Population \citep{eurostat2013}. Following the same notation as in Section \ref{sec:incidence} we considered the following scenarios:
\begin{itemize}
    \item[] \textbf{Scenario 1:} $K_1=0$, $K_2=0$, and $K_3=0$, where only the intercept $a_0$ is different from 0.
    \item[] \textbf{Scenario 2:} $K_1=6$, $K_2=3$, and $K_3=4$, as an extreme scenario where only 4 coefficient are different from 0 (the intercept and the corresponding highest powers of the polynomial).
    \item[] \textbf{Scenario 3:} $K_1=4$, $K_2=2$, and $K_3=3$, where all coefficients are different from 0.
\end{itemize}
We assumed that 30 years of incidence data were available, with an additional 10 years set aside for testing purposes. We used uniform distributions to simulate coefficients associated with the age and period powers, while we used Gaussian distributions to simulate coefficients of the cohort effects (see Equation \ref{eq:incidence}). Finally, we simulated incident cases from a Poisson distribution with parameter given by the product of the true incidence rates and the corresponding European standard population size. For each scenario, we simulated 250 datasets. The simulation was implemented in R version 4.4.3 \citep{rsoftware}, and the code is available at \href{https://github.com/fabradimitra/Lasso-driven_APC_model_selection_simulation}{https://github.com/fabradimitra/Lasso\-driven\_APC\_model\_selection\_simulation}.

We compared our proposed model selection algorithm with the standard forward stepwise procedure with LRT ($ \alpha = 5 \% $). The parameter $h_{max}$ was set to 5, 10, 15, and 20, and the maximum powers in the polynomial of Equation (\ref{eq:incidence}) were set to $K_1=7$, $K_2=7$, and $K_3=7$ for both the stepwise and the proposed procedure summarized in Algorithm \ref{alg:apcsel}. We labeled the selected models as \textit{lasso\_hmax5},...,\textit{lasso\_hmax20} and \textit{lrt} for the lasso-selected models and the stepwise-selected model, respectively. 

\begin{figure}
  \centering
  \includegraphics[scale=0.54]{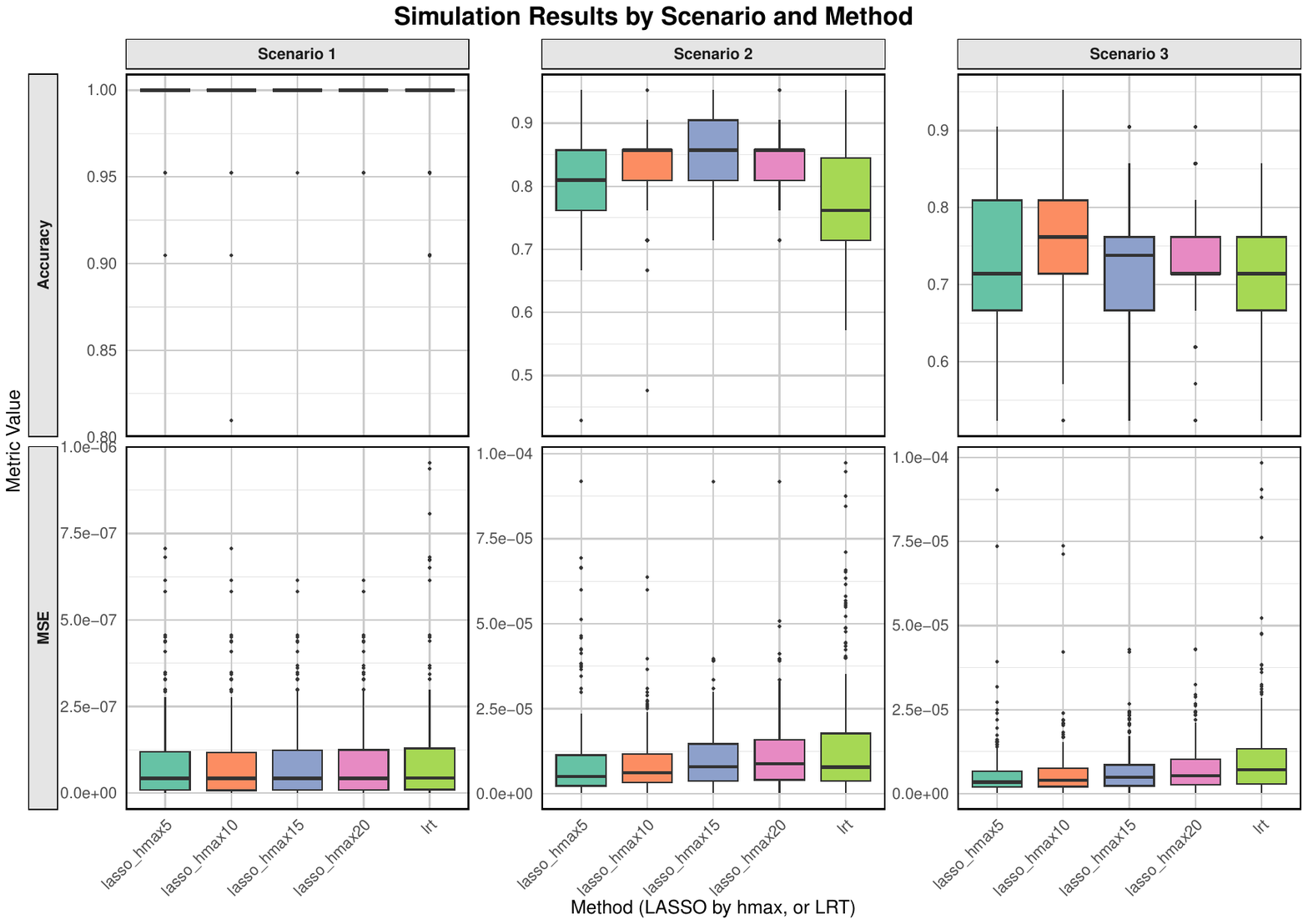} 
  \caption{Boxplots of the distributions of accuracy and MSE for each simulation scenario. The models were selected using Algorithm \ref{alg:apcsel} with different values of the hyper-parameter $h_{max}$ (\textit{lasso\_hmax5}, \ldots, \textit{lasso\_hmax20}) or a forward stepwise procedure based on the LRT with significance level $\alpha = 5\%$ (\textit{lrt}).}
  \label{fig:simulation}
\end{figure}

We considered two performance measures: the first measure is accuracy, namely total number of correctly selected variables and correctly discarded variables over the total number of variables; the second measure is MSE of the predicted incidence rates over the last 10 years, not involved in the selection process. Ideally, the first measure should be close to one, indicating that the selected model identifies, overall, more true variables and discards more incorrect variables from the true model. The MSE should instead be close to zero, indicating that the model generalizes well to unobserved incidence rates, which would then be used to achieve better predictions of prevalence. In Figure \ref{fig:simulation} we reported the distributions of the two performance measures for each selected model using boxplots.
In Scenario 1, where the incidence rate is constant, there is no difference in performance between the model selection algorithms considered. In Scenario 2 and 3, the models selected by our proposed algorithm achieve higher accuracy and lower MSE than those selected using the stepwise procedure. The distributions of the MSE seem to not be affected by the choice of $h_{max}$, except when we choose the hyper-parameter $h_{max}$ to be greater or equal than half of the years of the observation period, resulting in slightly worse MSE.

\section{Prevalence of colon cancer in Sweden}
\label{sec:appl}
The data considered in this application comes from two sources: the Swedish Cancer Registry, which provided incidence and follow-up data for patients diagnosed with colon cancer in Sweden between 1958 and 2019, with follow-up until December $31^{\text{st}}$, 2020, and the \citet{HMD}, which provided information on the Swedish population size up to 2024 and population mortality rates up to 2023. 

With the aim of providing accurate estimates of colon cancer prevalence from 1958 to 2019 and projections until 2030, we first compared the procedures to select the APC model described in Section \ref{sec:incidence}. In particular, we implemented our proposed LASSO-driven algorithm (see Algorithm \ref{alg:apcsel}) in R, setting $h_{max}$ equal to $20$ years, using the R package \textit{glmnet} \citep{friedman2010} for LASSO regularization in Generalized Linear Models. The forward stepwise procedure based on the LRT was implemented using the PIAMOD software \citep{deangelis1994}, following standard practice in population-based cancer registries. 

Stratifying the dataset by sex, for males we obtained a model with $K_1 = 7$, $K_2 = 0$, and $K_3 = 2$ using LRT, and a model with $K_1 = 6$, $K_2 = 2$, and $K_3 = 2$ using the LASSO-driven selection. For females, we obtained a model with $K_1 = 4$, $K_2 = 2$, and $K_3 = 7$ using LRT, and, as for males, a model with $K_1 = 6$, $K_2 = 2$, and $K_3 = 2$ using the LASSO-driven selection. For both sexes, the LASSO approach yielded a model with 7 degrees of freedom, as several intermediate powers of the APC polynomial were shrunk to 0. Plots of incident cases by calendar period and sex showed similar trends between the LRT and LASSO-driven selected model, despite the difference in the number of parameters between the models. Further details can be found in Figure \ref{fig:incidencetrend} of the Appendix, Section \ref{suppl:incidence}. 

We used four additional years of data on colon cancer incident cases provided by the Swedish National Board of Health and Welfare \citep{socialstyrelsen} to compare the generalization performance of the two models. Results are shown in Table \ref{tab:incident_cases_by_sex}. 
\begin{table}[!ht]
\centering
\caption{Observed colon cancer incident cases for the extra four years 2020--2023 provided by the Swedish National Board of Health and Welfare \citep{socialstyrelsen} and estimated incident cases by the two considered APC models with percentage relative errors (RE\%) by Sex.}
\small
\begin{tabularx}{\linewidth}{
    l 
    >{\centering\arraybackslash}X
    >{\centering\arraybackslash\hsize=1.4\hsize}X
    >{\centering\arraybackslash\hsize=0.8\hsize}X
    >{\centering\arraybackslash}X
    >{\centering\arraybackslash\hsize=0.8\hsize}X
}
\toprule
\textbf{Year} & \textbf{Observed Cases} & \textbf{LASSO Estimates (95\% CI)} & \textbf{RE\% (LASSO)} & \textbf{LRT Estimates} & \textbf{RE\% (LRT)} \\
\midrule
\textit{Males} & & & & & \\
\midrule
2020 & 2457 & 2591.89 (2548.26--2636.26) & 5.5 & 2491.11 & \textbf{1.4} \\
2021 & 2686 & 2666.13 (2617.92--2715.23) & \textbf{0.7} & 2562.24 & 4.6 \\
2022 & 2746 & 2793.26 (2687.88--2891.62) & \textbf{1.7} & 2632.85 & 4.1 \\
2023 & 2806 & 2810.70 (2754.60--2867.94) & \textbf{0.2} & 2702.20 & 3.7 \\
\midrule
\textit{Females} & & & & & \\
\midrule
2020 & 2591 & 2570.72 (2527.61--2614.56) & \textbf{0.8} & 2518.42 & 2.8 \\
2021 & 2747 & 2636.16 (2588.49--2684.71) & \textbf{4.0} & 2571.12 & 6.4 \\
2022 & 2738 & 2701.94 (2652.55--2752.25) & \textbf{1.3} & 2623.19 & 4.2 \\
2023 & 2870 & 2767.68 (2713.73--2822.71) & \textbf{3.6} & 2674.37 & 6.8 \\
\bottomrule
\end{tabularx}
\label{tab:incident_cases_by_sex}
\end{table}
We observe that, except for males in the calendar year 2020, the estimated incident cases from the LASSO-selected model are closer to the observed cases. Confidence intervals for the projections of the LASSO-selected model were calculated in R using the multivariate delta method \citep{casella2024}, whereas the PIAMOD software does not provide confidence intervals for projections of neither incidence nor prevalence.

Net survival estimation was carried out separately for different sex and age-at-diagnosis groups, which generally exhibit similar cancer survival characteristics. Specifically, we considered the following age groups: 0-49, 50-59, 60-69, 70-79, and 80 years and older. We compared two models: WMCM, proposed by \cite{lambert2007}, and FPSCM described in Section \ref{sec:netsurvival}. Each parameter of the models was specified as function of the covariates age at diagnosis, calendar year at diagnosis, and their interaction. The baseline excess hazard of the FPSCM was modelled using a restricted cubic spline function of the logarithm of time, with five degrees of freedom to ensure sufficient flexibility \citep{syriopoulou2019}. The cure models were implemented in R using the package \textit{cuRe} \citep{jensen2022} fitted on the individual follow-up data stratified as mentioned above.
\begin{table}[htbp]
\centering
\caption{Comparison of flexible parametric survival cure model (FPSCM) and weibull mixture cure model (WMCM) by AIC and BIC for different age groups and sex.}
\small
\begin{tabularx}{\linewidth}{
    l 
    >{\centering\arraybackslash}X
    >{\centering\arraybackslash}X
    >{\centering\arraybackslash}X
    >{\centering\arraybackslash}X
}
\toprule
 & \multicolumn{2}{c}{\textbf{AIC}} & \multicolumn{2}{c}{\textbf{BIC}} \\
\cmidrule(lr){2-3} \cmidrule(lr){4-5}
 & \textbf{FPSCM} & \textbf{WMCM} & \textbf{FPSCM} & \textbf{WMCM} \\
\midrule
\textit{Males} & & & & \\

0--49  & \textbf{18438.59} & 18616.66 & \textbf{18523.17} & 18694.73 \\
50--59 & \textbf{36929.71} & 37072.03 & \textbf{37022.12} & 37157.33 \\
60--69 & \textbf{84050.12} & 84539.72 & \textbf{84153.96} & 84635.57 \\
70--79 & \textbf{110771.5} & 112152.3 & \textbf{110880.4} & 112252.8 \\
80+    & \textbf{47945.77} & 49558.89 & \textbf{48048.27} & 49653.51 \\
\midrule
\textit{Females} & & & & \\

0--49  & \textbf{20266.26} & 20378.01 & \textbf{20352.34} & 20457.47 \\
50--59 & \textbf{38091.84} & 38268.16 & \textbf{38184.75} & 38353.92 \\
60--69 & \textbf{82559.37} & 82969.25 & \textbf{82662.42} & 83064.37 \\
70--79 & \textbf{117186.9} & 118189.0 & \textbf{117295.8} & 118289.6 \\
80+    & \textbf{67633.74} & 69165.31 & \textbf{67739.87} & 69263.28 \\
\bottomrule
\end{tabularx}
\label{tab:aic_bic_models}
\end{table}

Table \ref{tab:aic_bic_models} presents the Akaike Information Criterion (AIC) \citep{akaike1973} and the Bayesian Information Criterion (BIC) \citep{schwarz1978} for the two models considered, across different age groups and sexes. As shown in the table, the FPSCM provides the best trade-off between model fit and parsimony in all scenarios, since for each age-stratified model both the AIC and the BIC are smaller for the FPSCM than for the WMCM.

We also performed a graphical validation of the two models against the non-parametric Pohar-Perme estimator as suggested in \cite{dimari2026}. Figures \ref{fig:netfemales} and \ref{fig:netmales} in Appendix \ref{suppl:graphicalcure} show better agreement between the FPSCM and the Pohar-Perme estimator than between the WMCM and the Pohar-Perme estimator, particularly in the tail of the net survival curve, which plays a crucial role in survival extrapolation.

After having selected the best incidence and net survival estimation approaches, we are now ready to estimate complete colon cancer prevalence proportions in Sweden from 1958 to 2019, and then predict prevalence proportions up to 2030. 
\begin{figure}
  \centering
  \includegraphics[scale=0.55]{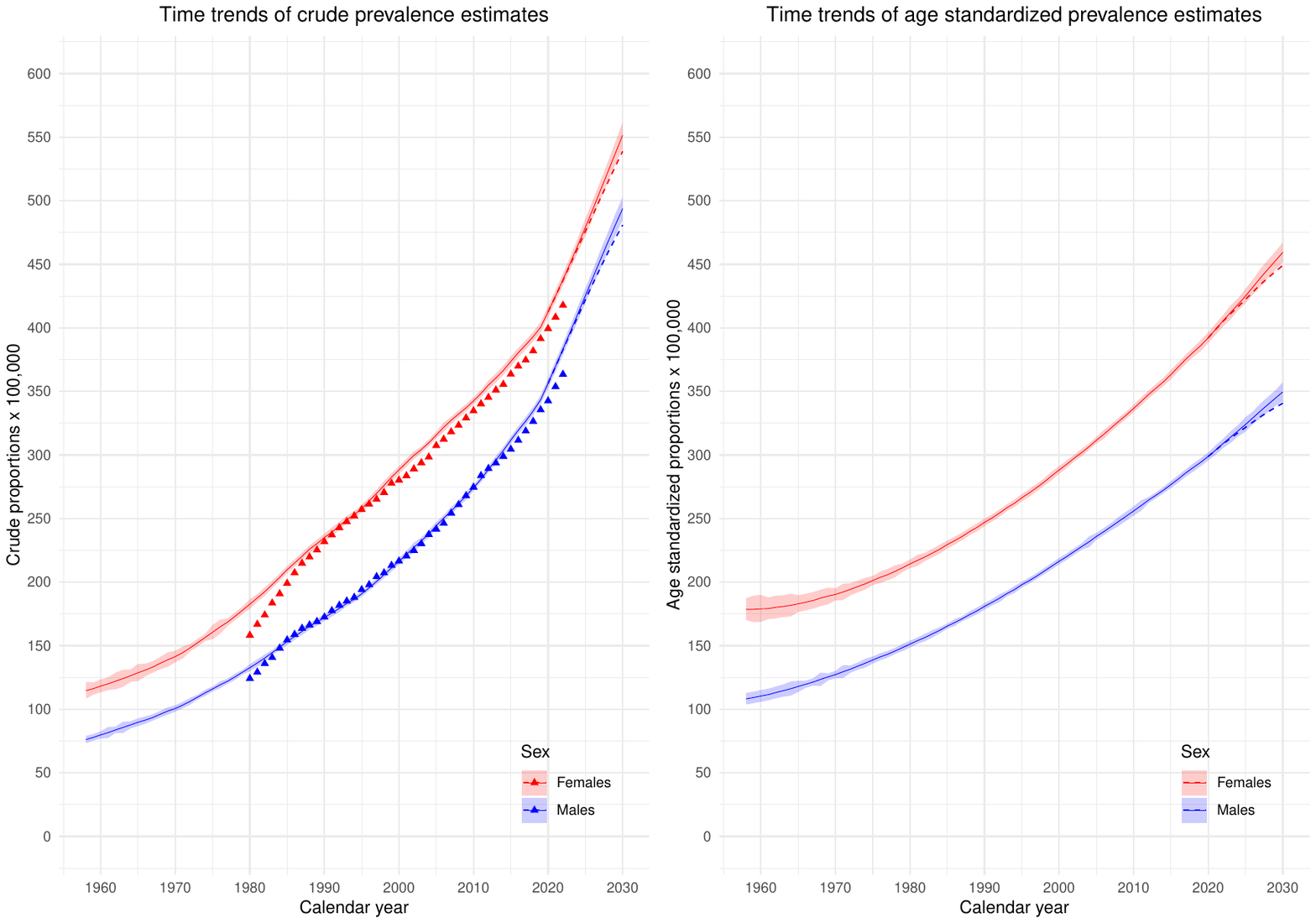} 
  \caption{Trends in colon cancer prevalence proportion per 100\,000 persons by calendar period and sex from 1958 to 2030. The left panel shows crude proportions, while the right panel displays age-standardized proportions using the Swedish population in 2000 as the standard. The triangles represent observed prevalence proportion (per 100\,000 persons) reported by NORDCAN \citep{nordcan2024}, while the lines show estimates from PIAMOD with the proposed pipeline (LASSO+FPSCM) and  95\% confidence intervals.}
  \label{fig:prevalencetrends}
\end{figure}
Figure \ref{fig:prevalencetrends} shows the estimated trends in colon cancer prevalence per 100\,000 persons in Sweden, obtained using the PIAMOD software with inputs from the proposed incidence and survival models, together with 95\% confidence intervals calculated using the multivariate delta method. The gradients of prevalence with respect to the parameters of the incidence model were approximated using finite differences. This calculation was very computationally demanding due to the number of execution of the PIAMOD software required to compute the centred difference quotient of prevalence with respect to each parameter of the incidence model. The variance-covariance matrix of the incidence model was instead obtained in R using the \textit{vcov} function. The left panel displays crude proportions, while the right panel presents age-standardized proportions using the Swedish population in 2000 as the standard, for females (in red), and males (in blue). We provided two projection scenarios: the solid lines represent estimates under the optimistic scenario, where the period trend of net survival is assumed to continue beyond the observed time period; the dashed lines represent projection estimates under the scenario where the period trend of net survival is assumed constant, equal to that of the last observed year, 2019. Throughout the study period, prevalence proportion increases and it is expected to increase also beyond the last observed year, for both sexes and both survival scenarios. The dots represent observed prevalence proportions (per 100\,000 persons) by sex provided by NORDCAN \citep{nordcan2024}. Although the NORDCAN estimates represent limited duration prevalence, the bias relative to prevalence is not expected to be large, given that the Swedish Cancer Registry began registration in 1958. The estimated prevalence proportions agreed well with the observed ones. We further observe that females have higher prevalence proportions than males for all calendar years considered. When we examine the age-standardized proportions with respect to the age distribution of the Swedish population in 2000 (plot on the right), we still observe an increase in prevalence proportions, but less steep than that observed for the crude proportions.
  
\section{Discussion}
\label{sec:conc}
We have proposed alternative modelling approaches in the PIAMOD framework \citep{verdecchia2002} to obtain more accurate estimates and predictions of prevalence. As we have shown in Section \ref{sec:model}, prevalence estimation requires a forward calculation procedure which involves estimating separately incidence from cancer diagnosis counts and net survival from incidence and follow-up registry data. The former requires the selection and estimation of an APC model (see Section \ref{sec:incidence}), and the latter a net survival model within a relative survival framework \citep{ederer1961}. 

Regarding the APC model, in this work we considered polynomials to account for trends in \textit{age}, \textit{period}, and \textit{cohort}, but regression splines \citep{durrleman1989} could also be used. Polynomial regression captures non-linear relationships using global basis functions, which makes it simple to implement and yields a smooth (infinitely differentiable), globally defined fit. However, higher-degree polynomials often suffer from numerical instability, high variance, and unrealistic boundary behaviours. Regression splines, in contrast, use piecewise polynomials with local support, allowing greater flexibility and improved stability. Their main drawbacks are increased model complexity and the need to select knot locations or smoothing parameters. Overall, polynomial regression can be effective for low-degree or globally smooth relationships, whereas regression splines are generally preferred for capturing complex, localized non-linear patterns, and if we want to impose specific boundary behaviours. The polynomial regression used in our case study on colon cancer of Section \ref{sec:appl} adapted very well to the observed data (see Figure \ref{fig:incidencetrend} of Appendix \ref{suppl:incidence}). Different cancer types or different diseases could instead require more flexibility and therefore the use of regression splines to capture the underlying APC trends.

For selecting the degrees of the APC model, we proposed using the LASSO path to build a set of solutions that would then be used to select the powers to include in the unpenalized model. We chose the powers associated to the lowest MSE on the incident cases over the last $h$ years, which we did not use during the estimation process. We then let $h$ vary between $1$ and an $h_{max}$ to build an ensemble of model selectors. The powers included in the polynomial of the APC model were the ones that appeared at least 50\% of the times across all model selectors. Finally, we used the model fitted to all available incidence years on the selected powers.

Regardless of the choice of the hyper-parameter $h_{max}$, the simulation study in Section \ref{sec:sim} showed that our proposed model selection algorithm yielded incidence models that better approximates the out-of-sample age-adjusted incidence rates. They also achieved higher selection accuracy of the polynomial powers in Equation (\ref{eq:incidence}) than the APC models obtained through the standard stepwise procedure based on the LRT. These results hold in both a sparse scenario (Scenario 2) and a full scenario (Scenario 3), with different degrees of the powers in the true model.

Regarding net survival estimation, we proposed using FPSCM within a relative survival framework \citep{jakobsen2020}, described in Section \ref{sec:netsurvival}. It is important to mention that the cure assumption does not necessarily need to be made. In the case of colon cancer, it is reasonable to assume that there is a fraction of individuals who will never experience the event of interest \citep{Di_Mari2025-xe}; however, when considering more severe cancers or other diseases, the cure assumption may not be plausible, and alternative approaches should be considered. Nevertheless, the PIAMOD framework allows the use of any net survival estimator, including models that do not assume a fraction of cured individuals.

The proposed method was applied on Swedish colon cancer patients diagnosed between 1958 and 2019, with follow-up until the end of 2020. Our LASSO-selected incidence model was more parsimonious than the one selected with the standard forward stepwise procedure and proved, on the one hand, to have comparable results in the observed years with respect to the less parsimonious model (see Figure \ref{fig:incidencetrend}), and, on the other hand, to have better generalization performance on extra incidence years provided by official statistics (see Table \ref{tab:incident_cases_by_sex}). Moreover, while we chose the hyper-parameter $h_{\text{max}}$ equal to 20, we performed an \emph{a posteriori} sensitivity analysis letting $h_{\text{max}}$ vary between 5 and 25 years in increments of 5 years, and the selected polynomial degrees remained the same for both males and females. What changed were the degrees of freedom. For males, the degrees of freedom remained at 6 up to $h_{\text{max}} = 15$ and increased to 7 for $h_{\text{max}} = 20$ and 25. For females, the pattern was reversed: the degrees of freedom were 7 from $h_{\text{max}} = 5$ to $h_{\text{max}} = 15$, and 6 for $h_{\text{max}} = 20$ and 25. As a result, the proposed model selection algorithm also proved to be not particularly sensitive to the choice of its hyper-parameter when applied to real data.

We then used model selection criteria to choose the best net survival model between the WMCM and FPSCM. For all age-stratified models, we proved how FPSCM had better AIC and BIC than WMCM. The better graphical adaptation to the non-parametric Pohar-Perme estimator \citep{perme2012}, particularly at the tail of the survival function, is shown in Figure \ref{fig:netfemales} and \ref{fig:netmales} of Appendix \ref{suppl:graphicalcure}.

Subsequently, we used the selected models to calculate colon cancer prevalence, and validated those estimates against the observed values provided by NORDCAN \citep{nordcan2024}. 

The cross-sectional reconstruction of prevalence proportion was accurate, with a slight overestimation in the earliest and most recent periods compared to NORDCAN prevalence observations. The overestimation in the early years was most likely due to the bias arising from the use of limited duration prevalence in NORDCAN to estimate prevalence, given that the  Swedish Cancer Registry was established in 1958. Patients diagnosed with colon cancer before that year who were still alive in the first years reported by NORDCAN were not included in the calculation. The slight overestimation in the most recent years could be due to reporting delays, since data from cancer registries are updated or revised regularly as new or corrected information is received \citep{laronningen2023}. The number of people with a past diagnosis of colon cancer was estimated to increase over time, and this trend is expected to continue in the future. One of the major drivers of the increasing prevalence proportions is population ageing. This can be observed from the age-standardized proportions, which also showed an increasing trend, albeit much less steep than the crude proportions. 
Furthermore, the proportion of female survivors is higher than that of males in both past and future years. Since colon cancer incidence rates are comparable between the sexes (see Figure \ref{fig:incidencetrend}), this reflects the generally better life expectancy of women compared to men, a phenomenon observed in the population as a whole, although the gap in Sweden has begun to narrow \citep{sundberg2018}.

From the illness-death model illustrated in Figure \ref{fig:compartimental}, it is also possible to reconstruct cross-sectional cause-specific mortality by combining prevalence, net survival, and incidence in a back-calculation procedure. Further details can be found in \citet{verdecchia1989}. Comparison with cause-specific mortality represents an additional external validation of incidence, survival and prevalence estimates, since mortality data were not used in the estimation process. The Swedish National Board of Health and Welfare \citep{socialstyrelsen} provides age-specific colon cancer mortality data from 1997 to 2023. 

\begin{figure}
  \centering
  \includegraphics[scale=0.54]{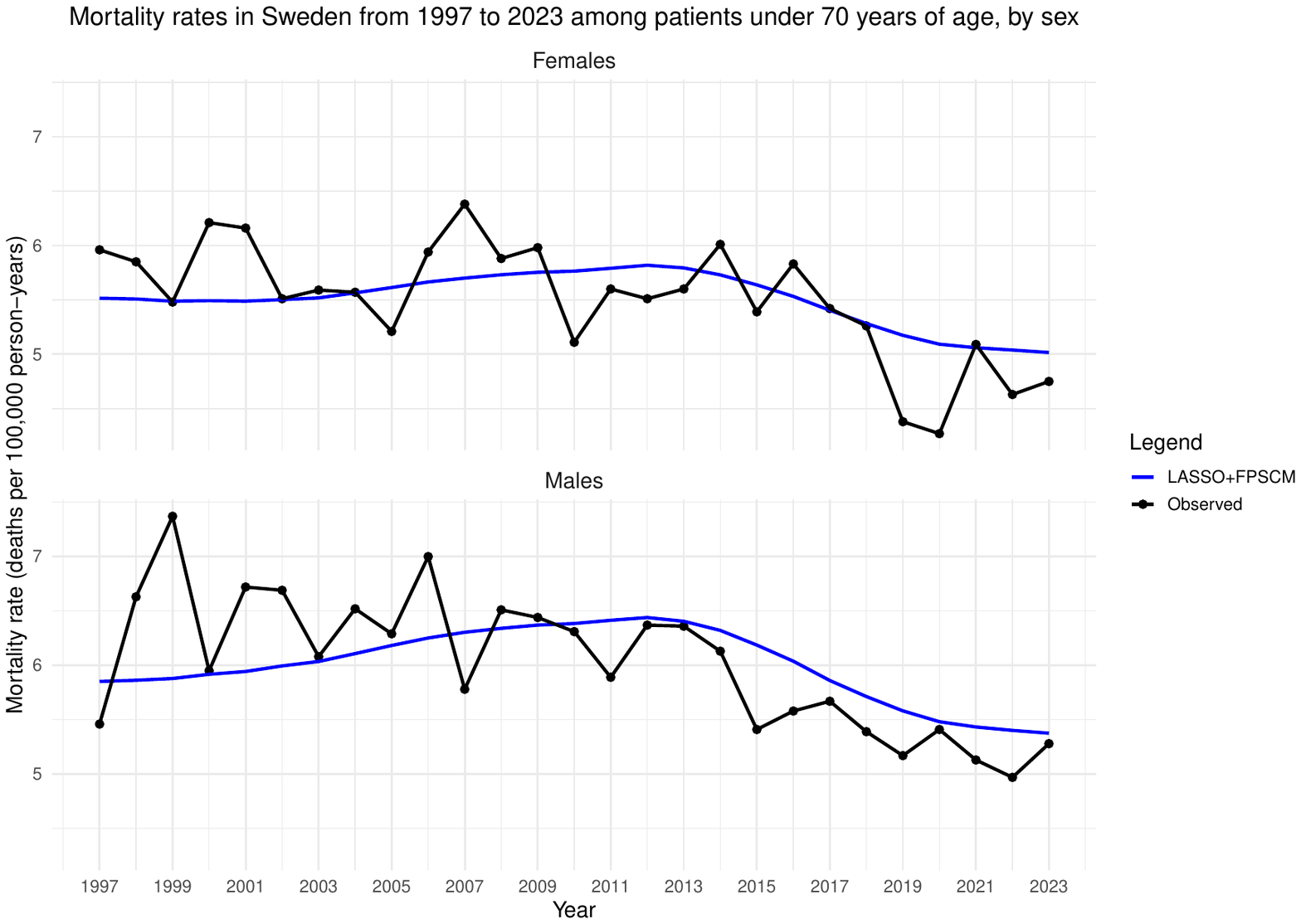} 
  \caption{Observed colon cancer mortality rates per 100,000 person-years in Sweden from 1997 to 2023 for females (upper panel) and males (lower panel), and colon cancer mortality estimated with our proposed pipeline in blue, the LASSO-driven APC model selection for incidence and a FPSCM for net survival (LASSO+FPSCM).}
  \label{fig:mortalitycomparison}
\end{figure}

In Figure \ref{fig:mortalitycomparison}, we compared the observed and estimated colon cancer mortality rates for patients under 70 years of age, given the potential over-reporting of cancer deaths among older patients \citep{mattsson1984, gerhardsson1990}. This over-reporting bias has also been observed more recently for prostate cancer \citep{orrason2021}, where the authors compared prostate cancer net survival estimated within the relative survival framework to that estimated using the cause-specific framework, finding that the two were more similar when patients aged 70 years and older were excluded. Similarly, we found that excluding patients over 70 years resulted in estimates that were much closer to those reported by the public source than when all age groups were considered. As shown in Figure \ref{fig:mortalitycomparison}, within this restricted age group, the cross-sectional reconstruction was accurate and aligned with the observed reduction in colon cancer mortality from 2010, most likely driven by the introduction of a population-based screening program in 2008 in the capital region of Stockholm for individuals aged 60 to 69 years (data not shown) \citep{petersson2020}. Mortality rates are expected to decrease further with the national colon cancer screening program, which has been in place since 2021, for individuals aged 60 to 74 years \citep{chaucastrand2024}.

Our proposed modelling approach also has limitations. First, the proposed APC model selection algorithm does not take sampling uncertainty into account, although we attempt to select more stable powers by constructing an ensemble of model selectors. Second, the confidence intervals for prevalence are overly optimistic, as they only account for the uncertainty arising from the APC model fitting while neglecting the uncertainty due to the net survival model. The latter source of uncertainty should not be a major concern, since population-based studies typically rely on large datasets. Nevertheless, accounting for all sources of uncertainty will be the subject of future research. Furthermore, the assumptions regarding the projected quantities, as described in Section \ref{sec:projections}, are stringent, and the current implementation of the PIAMOD software does not allow these assumptions to be relaxed \citep{deangelis1994}.

Future research could focus on the development of an R package for estimating and projecting prevalence, including additional incidence projection scenarios, additional net survival models and different population growth dynamics. Furthermore, smoothing spline tools \citep{wood2017} for the APC model can be applied, and compatibility with R packages such as \textit{rstpm2} or \textit{cuRe} \citep{liu2018,jensen2022} for net survival estimation can be ensured.

Meaningful analysis in population-based cancer studies requires models that capture the underlying data structure while remaining interpretable and generalizable. Our results show that up-to-date statistical methodology can improve complete cancer prevalence prediction while maintaining the framework simple and interpretable. Overall, our proposed approach has the potential to provide cancer registries with a useful tool for estimating and predicting complete cancer prevalence, thereby supporting cancer surveillance and more informed healthcare planning.

\section{Conflicts of interest}
The authors declare that they have no competing interests.

\section{Data availability}
The data used in this study contain individual patient information and are protected under the General Data Protection Regulation (GDPR). Consequently, they cannot be made publicly available. Aggregated data are publicly available from the relevant public authorities' websites mentioned in the main text.

\section{Author contributions statement}

F.D.M., R.D.A., R.R., and M.K. designed the methodology. F.D.M. conducted both the real-data application and the simulation study, and T.M.L.A., E.Y.-T.C., S.R., and P.W.D. analysed the results. F.D.M. wrote the manuscript, and all authors reviewed it.

\section{Acknowledgments}
The simulation study was performed thanks to the supercomputer Terastat 2: \\
\href{https://www.dss.uniroma1.it/it/HPCTerastat2}{https://www.dss.uniroma1.it/it/HPCTerastat2}.


\bibliographystyle{oup-abbrvnat}
\bibliography{reference}

@ARTICLE{capocaccia1997,
  title     = "Estimating the completeness of prevalence based on cancer
               registry data",
  author    = "Capocaccia, R and De Angelis, R",
  journal   = "Stat. Med.",
  publisher = "Wiley",
  volume    =  16,
  number    =  4,
  pages     = "425--440",
  month     =  feb,
  year      =  1997,
  language  = "en"
}

@ARTICLE{coldman1992,
  title     = "Calculating the prevalence of cancer",
  author    = "Coldman, A J and McBride, M L and Braun, T",
  journal   = "Stat. Med.",
  publisher = "Wiley",
  volume    =  11,
  number    =  12,
  pages     = "1579--1589",
  month     =  sep,
  year      =  1992,
  copyright = "http://onlinelibrary.wiley.com/termsAndConditions\#vor",
  language  = "en"
}

@ARTICLE{feldman1986,
  title     = "The prevalence of cancer",
  author    = "Feldman, Allen R and Kessler, Larry and Myers, Max H and
               Naughton, M Darlene",
  journal   = "N. Engl. J. Med.",
  publisher = "Massachusetts Medical Society",
  volume    =  315,
  number    =  22,
  pages     = "1394--1397",
  month     =  nov,
  year      =  1986,
  language  = "en"
}

@ARTICLE{deangelis2024,
  title     = "Complete cancer prevalence in Europe in 2020 by disease duration
               and country ({EUROCARE-6)}: a population-based study",
  author    = "De Angelis, Roberta and {Demuru, E. et al.} and {EUROCARE-6 Working Group}",
  journal   = "Lancet Oncol.",
  publisher = "Elsevier BV",
  volume    =  25,
  number    =  3,
  pages     = "293--307",
  month     =  mar,
  year      =  2024,
  language  = "en"
}

@ARTICLE{verdecchia1989,
  title     = "A method for the estimation of chronic disease morbidity and
               trends from mortality data",
  author    = "Verdecchia, A and Capocaccia, R and Egidi, V and Golini, A",
  journal   = "Stat. Med.",
  publisher = "Wiley",
  volume    =  8,
  number    =  2,
  pages     = "201--216",
  month     =  feb,
  year      =  1989,
  language  = "en"
}

@ARTICLE{mariotto2000,
  title     = "Using {AIDS} mortality data to reconstruct {HIV/AIDS} epidemics",
  author    = "Mariotto, A B and Verdecchia, A",
  journal   = "Stat. Med.",
  publisher = "Wiley",
  volume    =  19,
  number    =  2,
  pages     = "161--174",
  month     =  jan,
  year      =  2000,
  language  = "en"
}

@ARTICLE{verdecchia2002,
  title     = "Estimation and projections of cancer prevalence from cancer
               registry data",
  author    = "Verdecchia, Arduino and De Angelis, Giovanni and Capocaccia,
               Riccardo",
  journal   = "Stat. Med.",
  publisher = "Wiley",
  volume    =  21,
  number    =  22,
  pages     = "3511--3526",
  month     =  nov,
  year      =  2002,
  copyright = "http://onlinelibrary.wiley.com/termsAndConditions\#vor",
  language  = "en"
}

@ARTICLE{deangelis1994,
  title     = "{MIAMOD}: a computer package to estimate chronic disease
               morbidity using mortality and survival data",
  author    = "De Angelis, G and De Angelis, R and Frova, L and Verdecchia, A",
  journal   = "Comput. Methods Programs Biomed.",
  publisher = "Elsevier BV",
  volume    =  44,
  number    =  2,
  pages     = "99--107",
  month     =  aug,
  year      =  1994,
  language  = "en"
}

@ARTICLE{mariotto2006,
  title     = "Projecting the number of patients with colorectal carcinoma by
               phases of care in the {US}: 2000-2020",
  author    = "Mariotto, Angela B and Yabroff, K Robin and Feuer, Eric J and De
               Angelis, Roberta and Brown, Martin",
  journal   = "Cancer Causes Control",
  publisher = "Springer Science and Business Media LLC",
  volume    =  17,
  number    =  10,
  pages     = "1215--1226",
  month     =  dec,
  year      =  2006,
  language  = "en"
}

@ARTICLE{rossi2013,
  title    = "Estimates of cancer burden in Italy",
  author   = "Rossi, Silvia and Crocetti, Emanuele and Capocaccia, Riccardo and
              Gatta, Gemma and {AIRTUM Working Group}",
  journal  = "Tumori",
  volume   =  99,
  number   =  3,
  pages    = "416--424",
  month    =  may,
  year     =  2013,
  language = "en"
}

@ARTICLE{mariotto2020,
  title     = "Medical care costs associated with cancer survivorship in the
               United States",
  author    = "Mariotto, Angela B and Enewold, Lindsey and Zhao, Jingxuan and
               Zeruto, Christopher A and Yabroff, K Robin",
  journal   = "Cancer Epidemiol. Biomarkers Prev.",
  publisher = "American Association for Cancer Research (AACR)",
  volume    =  29,
  number    =  7,
  pages     = "1304--1312",
  month     =  jul,
  year      =  2020,
  language  = "en"
}

@ARTICLE{ederer1961,
  title    = "The relative survival rate: a statistical methodology",
  author   = "Ederer, F and Axtell, L M and Cutler, S J",
  journal  = "Natl. Cancer Inst. Monogr.",
  volume   =  6,
  pages    = "101--121",
  month    =  sep,
  year     =  1961,
  language = "en"
}

@ARTICLE{perme2012,
  title     = "On estimation in relative survival",
  author    = "Pohar Perme, Maja and Stare, Janez and Est{\`e}ve, Jacques",
  journal   = "Biometrics",
  publisher = "Wiley",
  volume    =  68,
  number    =  1,
  pages     = "113--120",
  month     =  mar,
  year      =  2012,
  copyright = "http://onlinelibrary.wiley.com/termsAndConditions\#vor",
  language  = "en"
}

@ARTICLE{deangelis1999,
  title     = "Mixture models for cancer survival analysis: application to
               population-based data with covariates",
  author    = "De Angelis, R and Capocaccia, R and Hakulinen, T and Soderman, B
               and Verdecchia, A",
  journal   = "Stat. Med.",
  publisher = "Wiley",
  volume    =  18,
  number    =  4,
  pages     = "441--454",
  month     =  feb,
  year      =  1999,
  language  = "en"
}

@ARTICLE{lambert2007,
  title     = "Estimating and modeling the cure fraction in population-based
               cancer survival analysis",
  author    = "Lambert, Paul C and Thompson, John R and Weston, Claire L and
               Dickman, Paul W",
  journal   = "Biostatistics",
  publisher = "Oxford University Press (OUP)",
  volume    =  8,
  number    =  3,
  pages     = "576--594",
  month     =  jul,
  year      =  2007,
  language  = "en"
}

@ARTICLE{hocking1976,
  title     = "A biometrics invited paper. The analysis and selection of
               variables in linear regression",
  author    = "Hocking, R R",
  journal   = "Biometrics",
  publisher = "JSTOR",
  volume    =  32,
  number    =  1,
  pages     = "1",
  month     =  mar,
  year      =  1976
}

@ARTICLE{novikov2020,
  title     = "A modified Prevalence Incidence Analysis Model method may
               improve disease prevalence prediction",
  author    = "Novikov, Ilya and Olmer, Liraz and Keinan-Boker, Lital and
               Silverman, Barbara and Robinson, Eliezer and Freedman, Laurence
               S",
  journal   = "J. Clin. Epidemiol.",
  publisher = "Elsevier BV",
  volume    =  123,
  pages     = "18--26",
  month     =  jul,
  year      =  2020,
  language  = "en"
}

@ARTICLE{lambert2015,
  title     = "Comparison of different approaches to estimating age
               standardized net survival",
  author    = "Lambert, Paul C and Dickman, Paul W and Rutherford, Mark J",
  journal   = "BMC Med. Res. Methodol.",
  publisher = "Springer Science and Business Media LLC",
  volume    =  15,
  number    =  1,
  pages     = "64",
  month     =  aug,
  year      =  2015,
  language  = "en"
}

@ARTICLE{andersson2011,
  title     = "Estimating and modelling cure in population-based cancer studies
               within the framework of flexible parametric survival models",
  author    = "Andersson, Therese M L and Dickman, Paul W and Eloranta, Sandra
               and Lambert, Paul C",
  journal   = "BMC Med. Res. Methodol.",
  publisher = "Springer Nature",
  volume    =  11,
  number    =  1,
  pages     = "96",
  month     =  jun,
  year      =  2011,
  language  = "en"
}

@ARTICLE{yu2013,
  title     = "Estimating the proportion cured of cancer: some practical advice
               for users",
  author    = "Yu, X Q and De Angelis, R and Andersson, T M L and Lambert, P C
               and O'Connell, D L and Dickman, P W",
  journal   = "Cancer Epidemiol.",
  publisher = "Elsevier BV",
  volume    =  37,
  number    =  6,
  pages     = "836--842",
  month     =  dec,
  year      =  2013,
  language  = "en"
}

@ARTICLE{dimari2023,
  title     = "{LASSO--penalized} clusterwise linear regression modelling: a
               two--step approach",
  author    = "Di Mari, Roberto and Rocci, Roberto and Gattone, Stefano Antonio",
  journal   = "J. Stat. Comput. Simul.",
  publisher = "Informa UK Limited",
  volume    =  93,
  number    =  18,
  pages     = "3235--3258",
  month     =  dec,
  year      =  2023,
  language  = "en"
}

@ARTICLE{schwarz1978,
  title     = "Estimating the dimension of a model",
  author    = "Schwarz, Gideon",
  journal   = "Ann. Stat.",
  publisher = "Institute of Mathematical Statistics",
  volume    =  6,
  number    =  2,
  pages     = "461--464",
  month     =  mar,
  year      =  1978
}

@ARTICLE{tibshirani1996,
  title     = "Regression shrinkage and selection via the lasso",
  author    = "Tibshirani, Robert",
  journal   = "J. R. Stat. Soc. Series B Stat. Methodol.",
  publisher = "Oxford University Press (OUP)",
  volume    =  58,
  number    =  1,
  pages     = "267--288",
  month     =  jan,
  year      =  1996,
  copyright = "https://academic.oup.com/journals/pages/open\_access/funder\_policies/chorus/standard\_publication\_model",
  language  = "en"
}

@ARTICLE{talback2011,
  title     = "Estimating expected survival probabilities for relative survival
               analysis--exploring the impact of including cancer patient
               mortality in the calculations",
  author    = "Talb{\"a}ck, Mats and Dickman, Paul W",
  journal   = "Eur. J. Cancer",
  publisher = "Elsevier BV",
  volume    =  47,
  number    =  17,
  pages     = "2626--2632",
  month     =  nov,
  year      =  2011,
  language  = "en"
}

@ARTICLE{park2007,
  title     = "\textit{L}1-regularization path algorithm for generalized linear
               models",
  author    = "Park, Mee Young and Hastie, Trevor",
  journal   = "J. R. Stat. Soc. Series B Stat. Methodol.",
  publisher = "Oxford University Press (OUP)",
  volume    =  69,
  number    =  4,
  pages     = "659--677",
  month     =  sep,
  year      =  2007,
  copyright = "https://academic.oup.com/journals/pages/open\_access/funder\_policies/chorus/standard\_publication\_model",
  language  = "en"
}

@ARTICLE{hakulinen1985,
  title     = "A computer program package for relative survival analysis",
  author    = "Hakulinen, T and Abeywickrama, K H",
  journal   = "Comput. Programs Biomed.",
  publisher = "Elsevier BV",
  volume    =  19,
  number    = "2-3",
  pages     = "197--207",
  year      =  1985,
  language  = "en"
}

@ARTICLE{jakobsen2020,
  title     = "Generalized parametric cure models for relative survival",
  author    = "Jakobsen, Lasse Hjort and B{\o}gsted, Martin and Clements, Mark",
  journal   = "Biom. J.",
  publisher = "Wiley",
  volume    =  62,
  number    =  4,
  pages     = "989--1011",
  month     =  jul,
  year      =  2020,
  copyright = "http://onlinelibrary.wiley.com/termsAndConditions\#vor",
  language  = "en"
}

@ARTICLE{jensen2022,
  title     = "Fitting parametric cure models in {R} using the packages cuRe
               and rstpm2",
  author    = "Jensen, Rasmus Kuhr and Clements, Mark and Gj{\ae}rde, Lars
               Klingen and Jakobsen, Lasse Hjort",
  journal   = "Comput. Methods Programs Biomed.",
  publisher = "Elsevier BV",
  volume    =  226,
  number    =  107125,
  pages     = "107125",
  month     =  nov,
  year      =  2022,
  copyright = "http://creativecommons.org/licenses/by/4.0/",
  language  = "en"
}

@ARTICLE{liu2018,
  title     = "Parametric and penalized generalized survival models",
  author    = "Liu, Xing-Rong and Pawitan, Yudi and Clements, Mark",
  journal   = "Stat. Methods Med. Res.",
  publisher = "SAGE Publications",
  volume    =  27,
  number    =  5,
  pages     = "1531--1546",
  month     =  may,
  year      =  2018,
  language  = "en"
}

@techreport{eurostat2013,
  title        = {Revision of the European Standard Population: Report of Eurostat’s Task Force},
  author       = {{Eurostat}},
  year         = {2013},
  institution  = {Publications Office of the European Union},
  address      = {Luxembourg},
  url          = {https://ec.europa.eu/eurostat/documents/3859598/5926869/KS-RA-13-028-EN.PDF}
}

@misc{socialstyrelsen,
  author = {{Socialstyrelsen}},
  title = {Cancer Incidence and Cause of Death Register in Sweden},
  year = {2026},
  note = {Swedish National Board of Health and Welfare}
}

@Manual{rsoftware,
  title = {R: A Language and Environment for Statistical Computing},
  author = {{R Core Team}},
  organization = {R Foundation for Statistical Computing},
  address = {Vienna, Austria},
  year = {2021},
  url = {https://www.R-project.org/},
}

@BOOK{casella2024,
  title     = "Statistical inference",
  author    = "Casella, George and Berger, Roger",
  publisher = "Chapman and Hall/CRC",
  month     =  apr,
  year      =  2024,
  address   = "Boca Raton",
  language  = "en"
}

@Inbook{akaike1973,
author="Akaike, Hirotogu",
editor="Kotz, Samuel
and Johnson, Norman L.",
title="Information Theory and an Extension of the Maximum Likelihood Principle",
bookTitle="Breakthroughs in Statistics: Foundations and Basic Theory",
year="1992",
publisher="Springer New York",
address="New York, NY",
pages="610--624",
isbn="978-1-4612-0919-5",
doi="10.1007/978-1-4612-0919-5_38",
}

@ARTICLE{orrason2021,
  title     = "Comparison of relative survival and cause-specific survival in
               men with Prostate Cancer according to age and risk category: A
               nationwide, population-based study",
  author    = "Orrason, Andri Wilberg and Garmo, Hans and Styrke, Johan and
               Dickman, Paul W and Stattin, P{\"a}r",
  journal   = "Am. J. Epidemiol.",
  publisher = "Oxford University Press (OUP)",
  volume    =  190,
  number    =  10,
  pages     = "2053--2063",
  month     =  oct,
  year      =  2021,
  copyright = "http://creativecommons.org/licenses/by/4.0/",
  language  = "en"
}

@misc{nordcan2024,
  author       = {Larønningen, S. and {Arvidsson, G. et al.}},
  title        = {{NORDCAN: Cancer Incidence, Mortality, Prevalence and Survival in the Nordic Countries, Version 9.4 (29.08.2024)}},
  year         = {2024},
  publisher    = {Association of the Nordic Cancer Registries. Cancer Registry of Norway},
  howpublished = {\url{https://nordcan.iarc.fr/}},
  note         = {Accessed on 09/05/2025}
}

@Article{friedman2010,
    title = {Regularization Paths for Generalized Linear Models via
      Coordinate Descent},
    author = {Jerome Friedman and Robert Tibshirani and Trevor Hastie},
    journal = {Journal of Statistical Software},
    year = {2010},
    volume = {33},
    number = {1},
    pages = {1--22},
    doi = {10.18637/jss.v033.i01},
  }

@ARTICLE{syriopoulou2019,
  title     = "Robustness of individual and marginal model-based estimates: A
               sensitivity analysis of flexible parametric models",
  author    = "Syriopoulou, Elisavet and Mozumder, Sarwar I and Rutherford,
               Mark J and Lambert, Paul C",
  journal   = "Cancer Epidemiol.",
  publisher = "Elsevier BV",
  volume    =  58,
  pages     = "17--24",
  month     =  feb,
  year      =  2019,
  copyright = "http://creativecommons.org/licenses/by/4.0/",
  language  = "en"
}

@ARTICLE{petersson2020,
  title     = "Increasing incidence of colorectal cancer among the younger
               population in Sweden",
  author    = "Petersson, J and Bock, D and Martling, A and Smedby, K E and
               Angenete, E and Saraste, D",
  journal   = "BJS Open",
  publisher = "Oxford University Press (OUP)",
  volume    =  4,
  number    =  4,
  pages     = "645--658",
  month     =  aug,
  year      =  2020,
  copyright = "http://creativecommons.org/licenses/by-nc-nd/4.0/",
  language  = "en"
}

@ARTICLE{chaucastrand2024,
  title     = "Impact of organised colorectal cancer screening on age-specific
               population incidences: evidence from a quasi-experimental study
               in Sweden",
  author    = "Chauca Strand, Gabriella and Str{\"o}mberg, Ulf and Forsberg,
               Anna and Bonander, Carl",
  journal   = "Eur. J. Epidemiol.",
  publisher = "Springer Science and Business Media LLC",
  volume    =  39,
  number    =  1,
  pages     = "87--96",
  month     =  jan,
  year      =  2024,
  copyright = "https://creativecommons.org/licenses/by/4.0",
  language  = "en"
}

@ARTICLE{laronningen2023,
  title    = "{Nordcan.R}: a new tool for federated analysis and quality
              assurance of cancer registry data",
  author   = "Lar{\o}nningen, Siri and {Skog, A. et al.}",
  journal  = "Front. Oncol.",
  volume   =  13,
  pages    = "1098342",
  month    =  aug,
  year     =  2023,
  language = "en"
}

@article{eloranta2020,
  title = {Cancer survival statistics for patients and healthcare professionals – a tutorial of real‐world data analysis},
  volume = {289},
  ISSN = {1365-2796},
  DOI = {10.1111/joim.13139},
  number = {1},
  journal = {Journal of Internal Medicine},
  publisher = {Wiley},
  author = {Eloranta,  S. and Smedby,  K. E. and Dickman,  P. W. and Andersson,  T. M.},
  year = {2020},
  month = jul,
  pages = {12–28}
}

@ARTICLE{linz1969,
  title     = "Numerial methods for Volterra integral equations with singular
               kernels",
  author    = "Linz, Peter",
  journal   = "SIAM J. Numer. Anal.",
  publisher = "Society for Industrial \& Applied Mathematics (SIAM)",
  volume    =  6,
  number    =  3,
  pages     = "365--374",
  month     =  sep,
  year      =  1969
}

@BOOK{zhou2025,
  title     = "Ensemble methods",
  author    = "Zhou, Zhi-Hua",
  publisher = "Chapman and Hall/CRC",
  month     =  jan,
  year      =  2025,
  address   = "Boca Raton",
  language  = "en"
}

@ARTICLE{gerhardsson1990,
  title     = "Colorectal cancer in Sweden. A descriptive epidemiologic study",
  author    = "Gerhardsson, M and Steineck, G and Norell, S E",
  journal   = "Acta Oncol.",
  publisher = "Informa UK Limited",
  volume    =  29,
  number    =  7,
  pages     = "855--861",
  year      =  1990,
  language  = "en"
}

@ARTICLE{mattsson1984,
  title     = "Completeness of the Swedish Cancer Register. Non-notified cancer
               cases recorded on death certificates in 1978",
  author    = "Mattsson, B and Wallgren, A",
  journal   = "Acta Radiol. Oncol.",
  publisher = "Informa UK Limited",
  volume    =  23,
  number    =  5,
  pages     = "305--313",
  year      =  1984,
  language  = "en"
}

@BOOK{wood2017,
  title     = "Generalized additive models",
  author    = "Wood, Simon N",
  publisher = "Chapman and Hall/CRC",
  month     =  may,
  year      =  2017
}

@article{morris2019,
  title = {Using simulation studies to evaluate statistical methods},
  volume = {38},
  ISSN = {1097-0258},
  DOI = {10.1002/sim.8086},
  number = {11},
  journal = {Statistics in Medicine},
  publisher = {Wiley},
  author = {Morris,  Tim P. and White,  Ian R. and Crowther,  Michael J.},
  year = {2019},
  month = jan,
  pages = {2074–2102}
}

@misc{HMD,
  author = {{Human Mortality Database}},
  title = {Human Mortality Database},
  institution = {University of California, Berkeley (USA), and Max Planck Institute for Demographic Research (Germany)},
  year = {2025},
  url = {https://www.mortality.org/},
  note = {Accessed: 2025}
}

@article{sundberg2018,
  title = {Why is the gender gap in life expectancy decreasing? The impact of age- and cause-specific mortality in Sweden 1997–2014},
  volume = {63},
  ISSN = {1661-8564},
  DOI = {10.1007/s00038-018-1097-3},
  number = {6},
  journal = {International Journal of Public Health},
  publisher = {Springer Science and Business Media LLC},
  author = {Sundberg,  Louise and Agahi,  Neda and Fritzell,  Johan and Fors,  Stefan},
  year = {2018},
  month = apr,
  pages = {673–681}
}

@article{durrleman1989,
author = {Durrleman, Sylvain and Simon, Richard},
title = {Flexible regression models with cubic splines},
journal = {Statistics in Medicine},
volume = {8},
number = {5},
pages = {551-561},
doi = {https://doi.org/10.1002/sim.4780080504},
year = {1989}
}

@article{dimari2026,
  title = {The relative survival framework: theory,  history,  and discussion},
  ISSN = {2281-695X},
  DOI = {10.1007/s40300-026-00312-6},
  journal = {METRON},
  publisher = {Springer Science and Business Media LLC},
  author = {Di Mari,  Fabrizio and Rocci,  Roberto and De Angelis,  Roberta},
  year = {2026},
  month = {June} 
}

@ARTICLE{Di_Mari2025-xe,
  title     = "A model-based approach to estimate the cure fraction in the
               population of colon cancer patients",
  author    = "Di Mari, Fabrizio and Rocci, Roberto and De Angelis, Roberta and
               Rossi, Silvia and Tagliabue, Giovanna",
  journal   = "Statistical Methods \& Applications.",
  publisher = "Springer Science and Business Media LLC",
  volume    =  34,
  number    =  5,
  pages     = "1057--1075",
  month     =  nov,
  year      =  2025,
  copyright = "https://creativecommons.org/licenses/by/4.0",
  language  = "en"
}


\begin{appendices}

\section{Incidence trends}
\label{suppl:incidence}
\begin{figure}[!ht]
  \centering
  \includegraphics[scale=0.54]{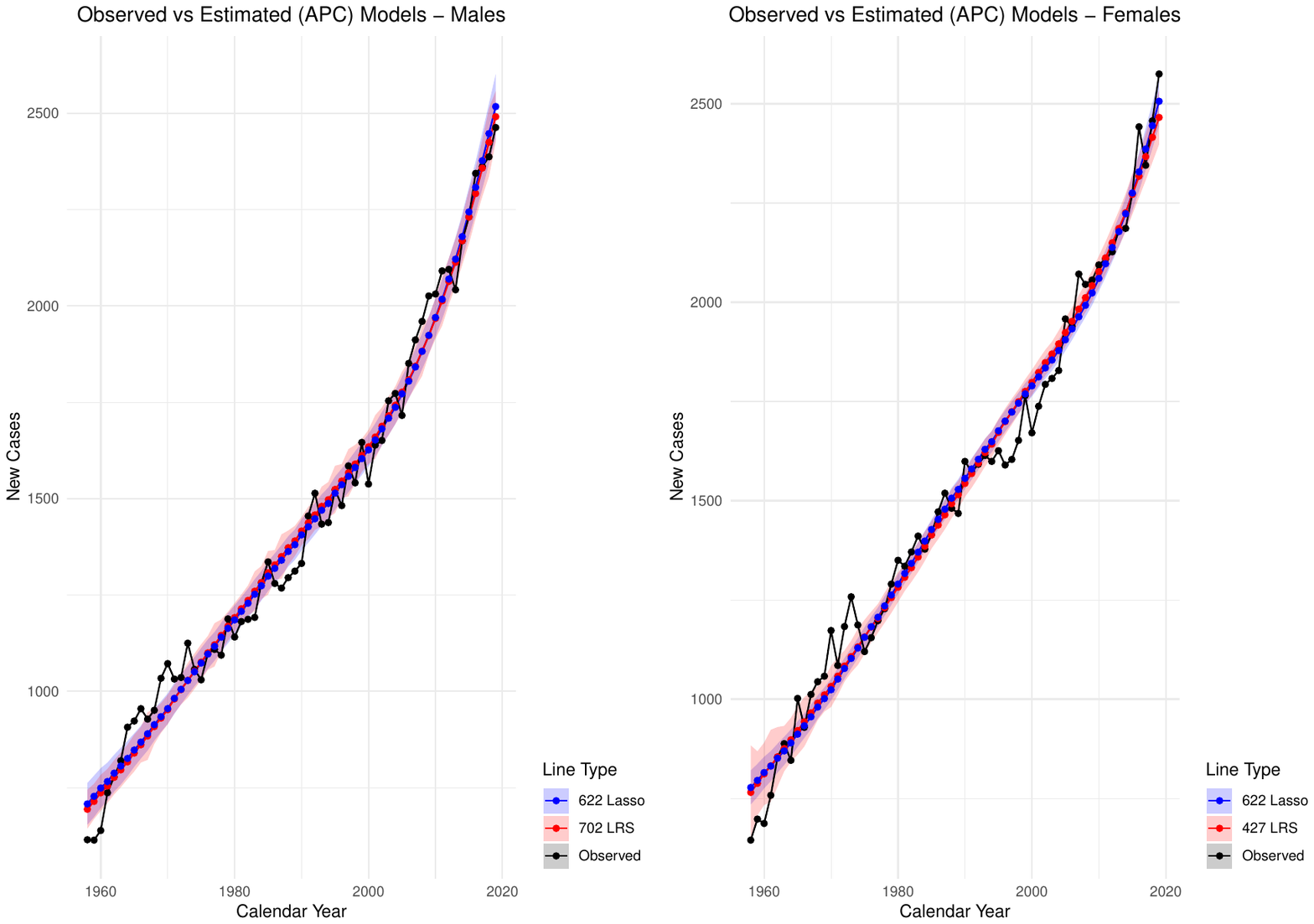} 
  \caption{Trends in colon cancer incident cases by calendar period and sex in the observation period 1958 to 2019. The black dots represents the observed incident cases, the red dots the incident cases estimated by the model selected through a forward stepwise procedure with LRT, and the blue dots the incident cases estimated by the model selected through the LASSO-driven procedure described in Section 2.1, with $h_{max}=20$ years. Data from the Swedish Cancer Registry.}
  \label{fig:incidencetrend}
\end{figure}

Trends in incident cases by calendar period and sex in the observation period are presented in Figure \ref{fig:incidencetrend}, with 95\% confidence intervals calculated using the multivariate delta method \citep{casella2024}.
The trends of the two models are fairly similar for both males and females, with a slightly more pronounced difference for females, where it is more evident that the LASSO-selected model attempts to locally adapt to the last observed years.

\section{Graphical comparison of the net survival models}
\label{suppl:graphicalcure}
\begin{figure}[!ht]
  \centering
  \includegraphics[scale=0.54]{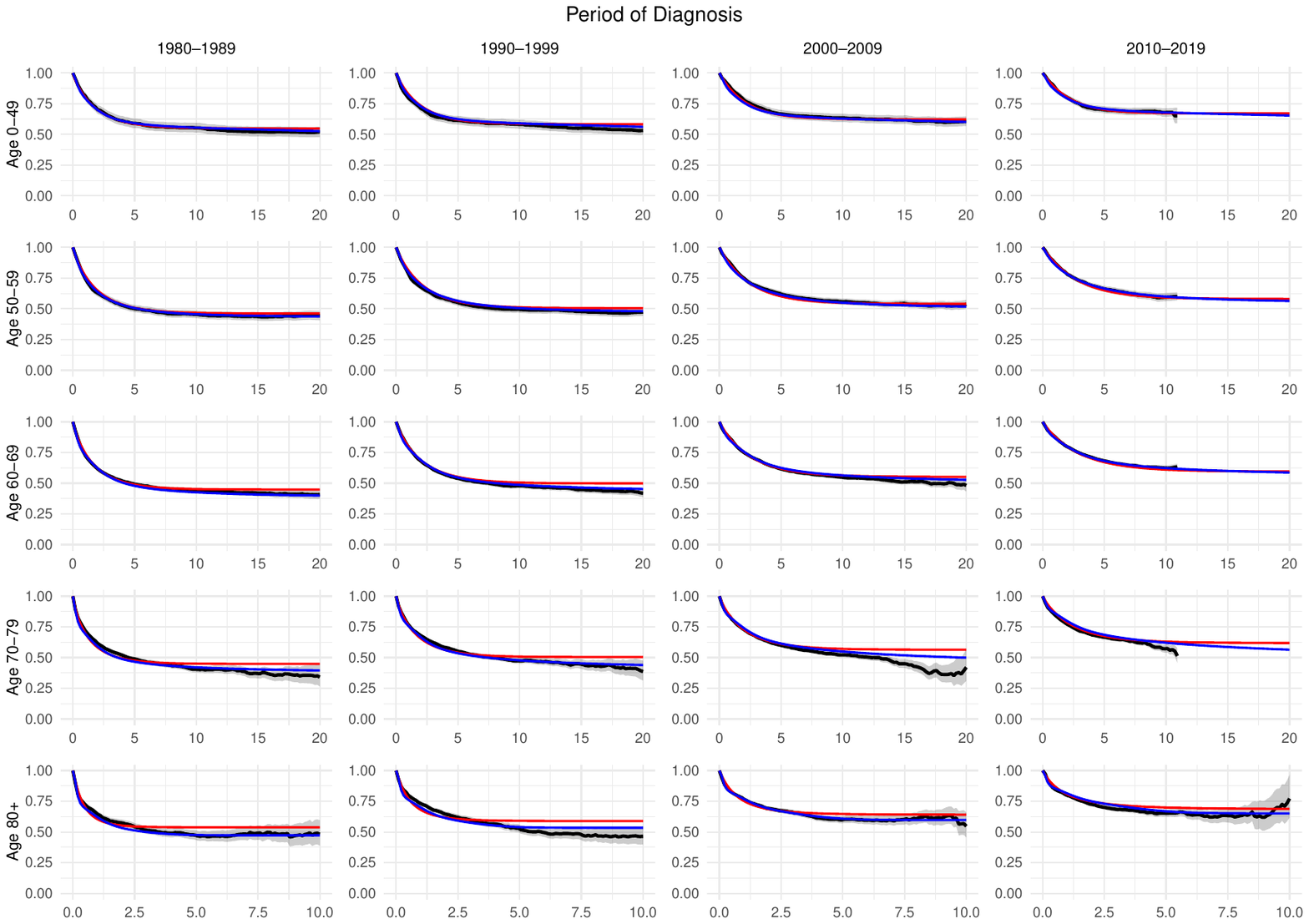} 
  \caption{Net survival of \textbf{male} colon cancer patients estimated by the Pohar-Perme estimator \citep{perme2012} (in black with a 95\% confidence intervals in gray), the WMCM (in red), and the FPSCM (in blue). The Y-axis represents the measured net survival probability, while the X-axis represents follow-up time in years.}
  \label{fig:netmales}
\end{figure}

\begin{figure}[!ht]
  \centering
  \includegraphics[scale=0.54]{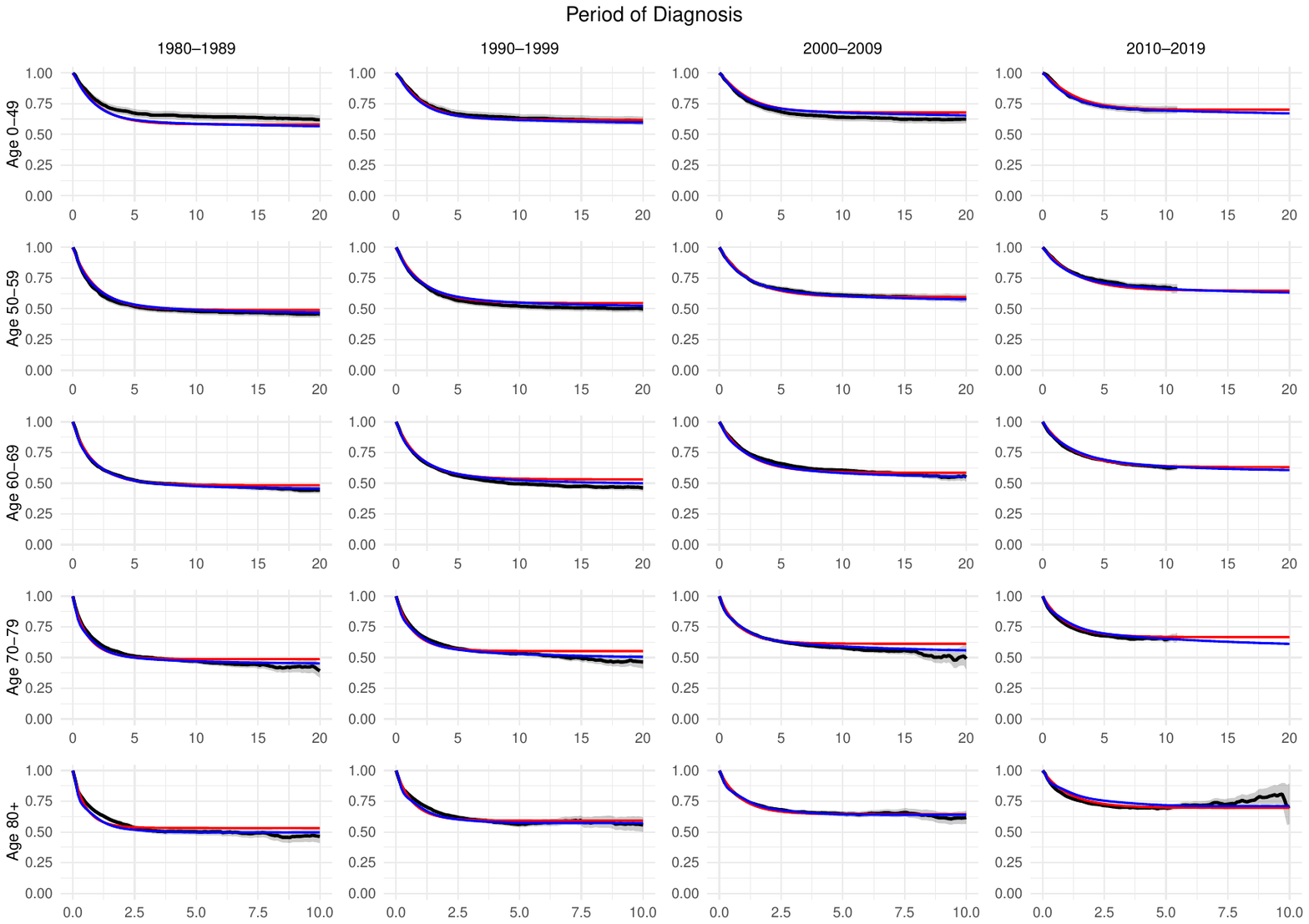} 
  \caption{Net survival of \textbf{female} colon cancer patients estimated by the Pohar-Perme estimator \citep{perme2012} (in black with a 95\% confidence intervals in gray), the WMCM (in red), and the FPSCM (in blue). The Y-axis represents the measured net survival probability, while the X-axis represents follow-up time in years.}
  \label{fig:netfemales}
\end{figure}

Figures \ref{fig:netmales} and \ref{fig:netfemales} show graphical comparisons of net survival estimates obtained using the Pohar-Perme estimator \citep{perme2012}, the WMCM \citep{lambert2007}, and the FPSCM \citep{jensen2022}. Each plot corresponds to a specific \textit{sex} and is organized by \textit{age at diagnosis}, divided into the categories 0-49, 50-59, 60-69, 70-79, and 80-99, and by \textit{period of diagnosis}, divided into four intervals: 1980-1989, 1990-1999, 2000-2009, and 2010-2019. Data from the 1958-1979 period are not shown.

\end{appendices}

\end{document}